# Closing the Evaluation Gap: Developing a Behavior-Oriented Framework for Assessing Virtual Teamwork Competency

Authors: Wenjie HU[1]* & Cecilia Ka Yuk Chan[1]

Affiliation[1]: The University of Hong Kong

Address: Teaching and Learning Innovation Center, Centennial Campus, The University of Hong Kong

*Corresponding author. Email: carol_hwj@connect.hku.hk

**Abstract:**

The growing reliance on remote work and digital collaboration has made virtual teamwork competencies essential for professional and academic success. However, the evaluation of such competencies remains a significant challenge. Existing assessment methods, predominantly based on self-reports and peer evaluations, often focus on short-term results or subjective perceptions rather than systematically examining observable teamwork behaviors. These limitations hinder the identification of specific areas for improvement and fail to support meaningful progress in skill development. Informed by group dynamic theory, this study developed a behavior-oriented framework for assessing virtual teamwork competencies among engineering students. Using focus group interviews combined with the Critical Incident Technique, the study identified three key dimensions—Group Task Dimension, Individual Task Dimension and Social Dimension—along with their behavioral indicators and student-perceived relationships between these components. The resulting framework provides a foundation for more effective assessment practices and supports the development of virtual teamwork competency essential for success in increasingly digital and globalized professional environments.

**Keywords:** 21st century abilities; Distance education and online learning; Evaluation methodologies; Post-secondary education; Cooperative/collaborative learning

## 1. Introduction

As online collaboration become integral to modern professional and academic environments, the ability to effectively engage in virtual teamwork has become a critical competency (Jony & Serradell-López, 2021; Koh et al., 2018). Effective virtual collaboration requires not only technical expertise but also behavioral and interpersonal competencies, such as communication, conflict resolution, and task coordination (Schulze & Krumm, 2017). However, research indicates that university students, particularly engineering students, often lack adequate virtual teamwork skills, facing challenges such as unequal task distribution, communication breakdowns, disengagement, and limited accountability, which undermine productivity and satisfaction (Ikonen et al., 2015). In response, higher education institutions have introduced various strategies, including workshops, structured activities, and hands-on projects, to help students improve their virtual teamwork competency (e.g. Chen et al., 2008).

Training in teamwork competency is inseparable from its evaluation, which is essential for identifying strengths, providing feedback, and guiding improvement (Varela & Mead, 2018). However, assessing virtual teamwork remains challenging, as teamwork is not easily quantified and lacks straightforward metrics for measurement (Britton et al., 2017). Current methods for assessing virtual teamwork skills largely rely on self-reports or peer evaluations (García et al., 2016a). However, these methods are prone to biases and have primarily focused on attitudinal and knowledge-based measures that capture learner outcomes over short-term periods (Chhabria et al., 2019). As a result, they often fail to accurately identify areas for growth or support meaningful progress.

In light of these challenges, this study seeks to develop a behavior-oriented framework for evaluating virtual teamwork competency, with a specific focus on engineering students. By identifying key behavioral indicators and their interrelationship, this framework is designed as a practical tool for educators and students, enabling effective assessment and fostering the enhancement of virtual teamwork competency.

## 2. Literature review

Virtual teamwork has become increasingly important for engineering students. Catalyzed by the COVID-19 pandemic, the prevalence of virtual instruction has risen notably (Wei et al., 2024). In professional contexts, remote collaboration has also become more common, as modern work environments demand effective collaboration in remote and globalized settings (Linnes, 2020). However, this shift brings a host of challenges that elevate the expectations for students' virtual teamwork competency. Key challenges include the absence of nonverbal cues, such as facial expressions and body language, which diminishes communication richness and complicates the interpretation of team members' contributions and intentions (Falls et al., 2014; Lepsinger & DeRosa, 2010). Technical limitations, including hardware, software, and network bandwidth issues, can disrupt collaborative processes and hinder the effectiveness of virtual tools designed to facilitate teamwork (Marra et al., 2016). Additionally, asynchronous communication, a common feature of virtual teams, can introduce delays, confusion, and a higher potential for conflict compared to face-to-face interactions (Vance et al., 2015).

Studies have shown that engineering students with strong virtual teamwork competency tend to achieve better project outcomes (e.g., Hosseini et al., 2018). However, many professionals report a lack of adequate training in virtual teamwork during their academic years, which leaves them unprepared for high-stakes virtual environments (Linnes, 2020; Brewer et al., 2015). As a result, academia has increasingly prioritized the development of virtual teamwork competency (e.g., Swartz et al., 2021). To maximize the effectiveness of this training, it is crucial to incorporate robust assessment practices. These practices, grounded in a comprehensive and structured theoretical framework, can provide the foundation for identifying key areas for improvement and tailoring development initiatives.

**2.1 Virtual teamwork competency assessment**

In virtual settings, traditional assessments often focus on outcomes, such as final deliverables or project quality, to gauge effectiveness. These assessments assume that the end product can reflect teamwork competency. However, this approach has significant limitations: Outcome-based assessments fail to capture teamwork processes, such as individual contributions, the quality of communication, or the division of labor (Vivian et al., 2016). Furthermore, group success may depend disproportionately on the efforts of a few individuals, obscuring whether every team member has acquired the necessary collaboration skills (Strom et al., 1999). As a result, this "black-box" approach often overlooks the interpersonal and procedural dynamics crucial to virtual teamwork, making it less effective for assessing true competency (Blanco et al., 2015). Recognizing these limitations, researchers have emphasized the importance of process-oriented assessments (Chan & Yeung, 2020). The most common method is self-evaluation using Likert scales, which may focus on participants' attitudes toward online teamwork, their perceptions of team effectiveness and on their own performance (Chavis et al., 2024; García et al., 2016b). While this approach is easy to implement and encourages students to reflect on their contributions, it has notable limitations. Firstly, the items used for self-assessment can often be vague, making it difficult to measure performance objectively (Chan, 2022; Fathi et al., 2019). Secondly, the subjectivity inherent in self-reports may lead to response distortions. Research indicates that students who are most confident in their teamwork abilities may actually possess the least accurate perceptions of their skills (Kotlyar et al., 2021; Zalesny, 1990). To address this issue, recent years have seen researchers turning to quantifiable data reflective of online environments to assess teamwork competency more objectively. For example, Tian et al. (2024) defined eight task-related member-level features analyze participants' team collaboration performance, which served as evidence to assess participants' virtual teamwork competency. Additionally, Petkovic et al. (2010) suggest utilizing collaborative tool usage statistics and time spent on project-related activities as indicators of teamwork activity. However, these methods may not effectively foster students' long-term teamwork competency, as they often fail to provide guidance on interpersonal interactions among team members. In response, some assessment methods have begun to focus on participants' interactions during collaboration. For instance, Guenaga et al. (2014) employed serious games for assessment, capturing user interaction data—such as help requests and communication with others—which provides insights into the collaborative process. However, there remains a lack of a comprehensive theoretical framework for assessing virtual teamwork competency based on behavior.

Behavior-based frameworks excel by focusing on observable interactions that drive collaboration. Traditional approaches centered on cognition and attitudes in teamwork can provide insights into individual knowledge and dispositions, yet they struggle to capture the dynamic, collective processes of team interactions (Aguado et al., 2014). Such methods frequently depend on self-reported data, introducing biases and varying perceptions that may not accurately reflect the team's collective viewpoint (Carron et al., 1985). In contrast, behavior-based frameworks emphasize the analysis of interactions such as communication, task coordination, and conflict resolution, offering a more objective understanding of team dynamics. According to Group Dynamics Theory, team performance emerges from the interplay of group interactions rather than individual efforts alone (Lewin, 1947). By linking specific teamwork behaviors to practical outcomes, these frameworks can help students enhance their understanding of how teams operate effectively and identify areas for improvement (Havyer et al., 2014).

**2.2 Theoretical framework: Group dynamics theory**

To guide the analysis and interpretation of the data, group dynamics theory was employed as a theoretical framework (Carron et al., 1985). By focusing on both individual and group-level behaviors, as well as task and social dimensions, this theory offers a holistic perspective that is well-suited for exploring the multifaceted nature of virtual teamwork.

At the individual level, the framework emphasizes the role of personal perceptions and behaviors in shaping group dynamics. Individual cohesion reflects the degree to which a member is attracted to the group and motivated to remain part of it, which directly influences their engagement and commitment (Carron et al., 1985). In contrast, the group level pertains to shared perceptions and collective interactions that bind the team together. Collective cohesion fosters a sense of "we-ness," where the group's identity and goals take precedence over individual interests, enhancing the team's ability to work as a cohesive unit (Forsyth, 2018). The framework further distinguishes between task-oriented cohesion and social-oriented cohesion. Task cohesion emphasizes the group's orientation toward achieving shared goals, ensuring focus and alignment. It is essential for keeping the group motivated to achieve its objectives (Beal et al., 2003) Social cohesion, on the other hand, highlights the development and maintenance of interpersonal relationships within the group. It encompasses the emotional bonds, camaraderie, and mutual respect that contribute to a positive group atmosphere. Additionally, the framework underscores the interplay between these dimensions. High task cohesion may lead to efficiency but can also result in burnout or strained relationships, while strong social cohesion without task cohesion might impede goal achievement. Striking a balance between individual and group-level contributions is also important in virtual settings, where the lack of face-to-face interaction can lead to misunderstandings or disengagement.

**2.2 Behavioral indicators of virtual teamwork competency**

Although there is currently no comprehensive behavior-based framework specifically designed for assessing virtual teamwork competency, existing frameworks consistently emphasize behavioral indicators, positioning them at the center of virtual teamwork models. For instance, Lisa and Tobin (2001) highlighted five critical intrateam processes—communication, decision-making, conflict management, leadership, and coordination—that are essential for virtual team functioning and effectiveness. Similarly, Siebdrat et al. (2009) observed that high-performing virtual teams excel in behaviors such as team effort, coordination, and mutual task support. These components encompass multiple dimensions of social and task-related behaviors, as well as individual and collective aspects.

Bringing these frameworks together, task-oriented behaviors at the individual level include self-management, adaptability, and initiative (Hertel et al., 2006). For instance, Gieure et al. (2022) highlight task completion as a critical indicator for evaluating individual task-oriented capabilities. At the group level, these behaviors encompass goal clarification, decision-making, and workflow synchronization (Mayer et al., 2023). Communication stands out as a key indicator, as the limitations of digital tools often lead to misunderstandings and uneven information distribution (Aritz et al., 2018). To mitigate these issues, team members should engage in structured and precise communication practices, such as summarizing key points and confirming meanings to ensure clarity (Piccoli et al., 2004). On the social-oriented dimension, Lacher & Biehl (2019) note that individuals with high social sensitivity can better facilitate team communication and performance in asynchronous text environments. Consequently, the importance of relationship-building behaviors has been emphasized by scholars (Hu & Chan, 2024; Mayer et al., 2023). Specific behaviors in this context include fostering mutual understanding, establishing and maintaining trust, and resolving conflicts (Willox et al., 2022; Yoon & Seung-Won, 2003).

**2.3 Behavioral indicator interplay**

As team dynamics represent a complex network of interconnections, where each member relies on others for critical information and the successful execution of shared tasks, teamwork behaviors are not isolated but are deeply interconnected, with each behavior influencing and reinforcing others in a complex interplay that shapes the collaborative experience (Avry et al., 2020). These cooperative interactions influence communication and, consequently, team performance by enabling or obstructing information sharing (Dickinson & McIntyre, 1997).

To gain a deeper understanding of virtual teamwork competency, it is essential to explore the interactions between different indicators. For example, performance monitoring and intrateam coaching are closely linked, with the former gathering data on team dynamics and the latter using this information to guide improvement through feedback (Rousseau et al., 2006). Communication also acts as the vital link that facilitates the transition from evaluating team members' performance to providing constructive feedback (McIntyre et al., 1988). Backup behavior, another critical indicator, involves anticipating team members' needs and balancing workloads under pressure (Porter et al., 2003). For backup behavior to be effective, it must build upon either prior performance monitoring or the proactive seeking of assistance, which can then facilitate targeted and timely intervention (Brannick et al., 1997). Moreover, the perception of performance monitoring can make or break its intended outcome. If team members regard this monitoring as overly rigorous supervision, it may create a sense of distrust, undermining any potential benefits and adversely affecting team morale (Salas et al., 2009).

Despite numerous studies highlighting the importance of specific behavioral indicators, there remains a significant research gap. First, there is no comprehensive behavior-based framework designed specifically for assessing virtual teamwork competency. Second, few studies have explored the interactions between these behavioral indicators, which are closely linked to teamwork effectiveness and should be included in virtual teamwork competency framework. This lack of exploration limits our understanding of how these behavioral indicators work together to drive the success of virtual teams. In light of these challenges, this study aims to develop a behavior-oriented framework to evaluate virtual teamwork competency in engineering students, focusing on key behavioral indicators and their interrelationships to support effective assessment and skill development. Research questions are as follows:

1. What are the key behavioral indicators for virtual teamwork competency in engineering students?
2. What are the interrelationships between the identified behavioral indicators?

## 3. Methodology

This study aims to develop a behavior-oriented framework of virtual teamwork competency for engineering students by employing a qualitative approach. Focus group interviews are conducted to identify specific behavioral indicators relevant to virtual teamwork and to explore their interrelationships.

### 3.1 Focus Group Interviews

Following the literature review, focus group interviews were conducted to gather qualitative data from third- and fourth-year undergraduate engineering students with prior experience in virtual team-based learning environments. Participants were recruited using a purposive sampling strategy at a science and engineering university in Northern China. Announcements were posted on the university's online platform to invite junior and senior engineering students to participate. The study required participants to be third- or fourth-year students and to have experience with online collaboration, specifically having completed at least one full team task online. Compared to freshmen, upper-year students had gained more online collaboration experience during the pandemic. Finally, a total of 40 students voluntarily joined the study, including 32 males and 8 females. Among them, 18 were third-year students, and

22 were fourth-year students. Each focus group consisted of 4-5 participants. The participants primarily came from the following disciplines: Electronic Information Engineering, Electrical Engineering and Automation, Mechanical Manufacturing and Automation, and Optoelectronic Information Engineering.

The interviews were semi-structured, combining open-ended questions with guided prompts to encourage participants to share their experiences while maintaining focus on the research objectives. To align with the study's focus on virtual teamwork, the interviews were conducted online, simulating the digital environments where participants' teamwork experiences occurred. By adopting the Critical Incident Technique, which pinpoints significant events affecting the outcome of a process (Flanagan, 1954), participants detailed incidents that exemplified particularly effective or ineffective teamwork. For example, participants were asked to "share an example of a highly effective team collaboration experience in an online setting" and reflect on "what actions or behaviors of team members contributed to its success." They were also prompted to "describe an instance where virtual team collaboration did not meet expectations" and consider "what caused this outcome." These open-ended questions allowed participants to provide detailed narratives of their experiences, highlighting specific behaviors, challenges, and strategies that influenced the outcomes of virtual teamwork. With participants' permission, all sessions were audio-recorded and transcribed for further analysis.

## 3.2 Data Analysis

The transcribed data were analyzed using a systematic coding process grounded in Group Dynamics Theory to identify key behavioral indicators of virtual teamwork competencies and their interrelationships. Initially, theory-driven coding segmented the data based on predefined constructs from Group Dynamics Theory. These codes were then refined through axial coding to uncover relationships between concepts, providing a deeper understanding of teamwork competencies. Finally, selective coding synthesized the findings into a cohesive framework highlighting the core behavioral indicators and their interrelationships. Throughout the analysis, constant comparison was employed to ensure consistency and rigor. Researchers also documented their reflections on potential biases to maintain transparency and objectivity.

## 4. Findings

Following the data analysis, we developed a behavior-oriented theoretical framework grounded in group dynamics theory to assess the virtual teamwork competency of engineering students. As illustrated in Figure 1, this framework comprises 15 behavioral indicators categorized into three dimensions: Group task dimension, Social dimension, Individual task dimension, and, as well as the interrelationships among these behavioral indicators.

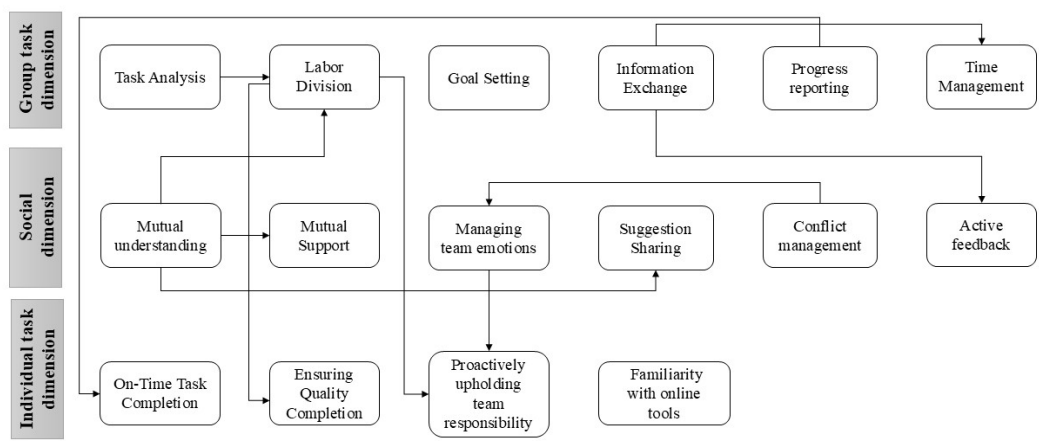

Figure 1 Behavior-oriented framework for assessing virtual teamwork competency

### 4.1 Group Task Dimension

In this dimension, which focuses on the systematic organization and execution of tasks within a team, **Task analysis** emerges as a critical component of online team collaboration. Task analysis involves breaking down projects into manageable parts and evaluating required resources and skills, which helps teams anticipate challenges and develop strategies. A student from Group 2 shared their successful experience in a competition, noting, "First, we analyze and dissect the competition situation. Based on the situation, we categorize the knowledge involved in the competition." Another group member added, "When working on a competition, we definitely anticipate some difficulties. We make an estimate and have some grasp of the challenges involved." This initial analysis helps students anticipate challenges and develop strategies to address them.

Moreover, this foundational step lays the groundwork for **goal setting**, which aligns team members toward a common objective. In online settings, where physical separation can hinder team cohesion, clear goal setting can align their efforts and maintain motivation throughout the project. For instance, a participant from Group 2 reflected on their experience, stating that "we had a strong determination to do well from the beginning. Even when facing significant challenges, we reminded ourselves of our initial purpose." Additionally, breaking down overarching goals into smaller tasks in virtual environments provides a clear pathway for progress. As a student from Group 7 explained, "when we divide a big goal into smaller ones, achieving each one brings a sense of accomplishment, which motivates us to continue." By setting clear goals and dividing them into manageable steps, teams can navigate challenges effectively while maintaining a shared sense of direction and motivation.

**Labor division** complements these efforts by ensuring tasks are distributed efficiently, enhancing team efficiency. Clear task allocation and active participation in task division are necessary to align efforts with shared goals. As one participant mentioned, "To improve efficiency online, it's essential to have clear task allocation and deadlines." Participants also emphasized that online collaboration requires members to be "proactive in understanding and assisting with task division," which ensures effective task allocation and alignment towards common goals. Following labor division,

**Information exchange** and **progress reporting** are crucial elements of online collaboration. In virtual settings, effective communication is essential for enhancing individual contributions. As one participant noted, "if everyone communicates more, it will definitely help improve their own part of the work. But if no one communicates, it feels like working in isolation, which is very limiting." Another participant emphasized the importance of increased communication in online collaboration, stating, "you have to compensate for the lack of face-to-face interaction by communicating more." Through regular communication, team members can stay connected and aligned after tasks are assigned. Progress reporting plays a vital role in this dynamic, as it involves sharing updates about tasks and achievements. This practice helps reduce uncertainty and fosters transparency within the team. As one participant highlighted, "the most important thing for online team collaboration is to provide real-time feedback on your progress, problems, or information." Additionally, sharing achievements can inspire and motivate the team, as noted by another participant: "showcasing your current progress and achievements, even if they seem small to others, can be encouraging to your team members." Overall, these practices help create a more engaged and cohesive team environment.

Similarly, **time management** is vital for coordinating schedules across different time zones and personal commitments. One participant mentioned that "our communication was limited" due to unfamiliarity and the lack of face-to-face interaction. To address this, strategies like sharing schedules and planning time slots that work for everyone were suggested. As another participant pointed out, "online communication is not very convenient because you have to consider what time is suitable to send messages." By implementing these strategies, teams can improve coordination, reduce delays, and enhance overall efficiency in their online collaborations.

**4.2 Social dimension**

The social dimension focuses on creating a supportive and cohesive team environment. One key behavioral indicator within this dimension is **Mutual understanding**. In the context of online collaboration, the absence of face-to-face interaction often exacerbates the sense of unfamiliarity among team members, as highlighted by one participant: "With my first online team, we were strangers and not familiar with each other, which led to many problems due to inadequate communication." To bridge this gap, participants suggested increasing communication beyond formal project discussions to include casual conversations that enhance team rapport. One participant advised, "Engage in more frequent communication, not just about work or collaboration, but also casual chats to build a connection." Additionally, having a team member who is "funny" can "create a relaxed atmosphere, encouraging more friend-like interactions." Building on mutual understanding, **Mutual support** also emerges as a pivotal behavior indicator that strengthens team bonds. When team members focus on their tasks while remaining ready to assist others, it fosters a harmonious team environment. As one participant observed, "If every online team member completes their tasks before helping others, the team atmosphere will be very harmonious." The online environment further encourages a culture of asking for help without hesitation, as noted by a participant who advised, "If you can't solve a problem on your own, you should definitely ask for help." This open exchange not only facilitates collaboration but also strengthens the team's dynamics.

Similarly, just as mutual support enhances teamwork and communication, **managing team emotions** also helps foster a positive and productive online environment. Participants emphasized that a harmonious atmosphere promotes better communication, while stress impedes it: "I feel that a harmonious atmosphere promotes communication, while a stressful one hinders it." To maintain this balance, strategies such as encouragement and humor are invaluable. One participant shared, "If we're close to a deadline and things aren't ready, I would motivate everyone in the group chat to push through." Another highlighted the role of humor, stating, "I try to introduce humor to create common topics for discussion, which can lighten the mood."

Participants also highlighted the importance of **suggestion sharing** in boosting productivity in online settings, though they acknowledged that it can be challenging. As one participant noted, "some teammates may not be as comfortable expressing themselves online," which can lead to fewer opinions and suggestions being shared. Despite these challenges, constructive suggestions play a crucial role in enhancing both individual effort and team performance. One participant stressed that "in online team collaboration, when everyone actively shares their ideas and discusses them together, it creates a supportive environment." Another participant illustrated this by describing how their partner's proactive suggestions significantly improved an online training session: "My partner approached the livestream from a student's perspective, identifying issues and proactively communicating them to me, which greatly improved the overall quality of the session." These examples underscore the critical role of suggestion sharing in fostering effective online collaboration.

**Conflict management** further strengthens the foundation of a cohesive and productive online team environment. Unresolved disputes can significantly hinder team progress, as highlighted by a participant from Group 3, who described a situation where conflicts arose due to inflexible attitudes: "four teammates argued frequently during online collaboration, and it became very unpleasant. Each person thought they were right and refused to compromise." Similarly, a participant from Group 5 emphasized how unresolved disagreements delayed their project, noting that "the software team member and I had different ideas, and our disagreement caused a significant delay in the team's mid-term progress." These examples illustrate the detrimental effects of poorly managed conflicts on team dynamics and efficiency. Conversely, successful teams often attribute their achievements to effective conflict resolution. As another participant from Group 3 explained, "while everyone had their own opinions, we coordinated our views, avoided arguments, and tried to solve problems peacefully." Appropriate conflict management can minimize disruptions while fostering a collaborative environment where diverse perspectives are integrated constructively.

Finally, **Active feedback** is another essential indicator of effective online collaboration, especially regarding interpersonal dynamics. It refers to timely and positively responding to team members' ideas, contributions, or questions, fostering an engaging and supportive environment. In virtual settings, the absence of feedback or dismissive responses can lead to frustration and a sense of disconnection. One participant shared, "Sometimes it feels like I'm the only one talking, and no one else responds—it gets really awkward." Another noted, "When I send a message online, some people don't reply at all, or they respond in a very cold, dismissive way. It makes the atmosphere feel unpleasant." Such behaviors can hinder communication flow and reduce motivation within the team.

**4.3 Individual task dimension**

In online team collaborations, self-management emerges as another critical dimension, with **On-time task completion** standing out as a key behavioral indicator in virtual environments where supervision is limited or absent. Participants often emphasized the difficulties of meeting deadlines in such settings. For example, one participant shared their experience during an online competition, noting that "everyone tends to have severe procrastination issues," which directly impacts their ability to meet deadlines. Another participant described the tendency to "slack off when working from home," noting that "without other team members to provide oversight, it requires greater self-discipline." These insights highlight the necessity for individuals to maintain personal accountability and effectively manage their time to ensure tasks are completed on schedule in online collaborations.

Equally important as timely completion is **Ensuring quality completion**. In virtual environments, team members may struggle to fully engage with their responsibilities. As one participant noted, "tasks might be marked as completed, but the quality often fell short." However, this challenge also highlights a key factor for successful online collaboration. Another participant shared a positive experience during an online competition, stating that "everyone had their own tasks and worked hard to do their part well." These contrasting examples emphasize the critical role of each team member's commitment to maintaining quality in online collaboration.

**Proactively upholding team responsibility** is another layer that enhances task quality and punctuality. Participants shared examples of how individual initiative can positively impact team dynamics. One participant recounted the story of a team member who, despite being unable to assist with physical tasks due to remote circumstances, actively sought ways to contribute. This individual "would proactively contact us, asking for tasks he could help with online," demonstrating a commendable willingness to engage and support the team. Such proactive behavior not only boosts team morale but also sets a positive example for others. As another participant noted, seeing someone "actively making contributions" can inspire others to step up their efforts, creating a ripple effect of increased enthusiasm and participation.

Finally, **Familiarity with online tools** is a crucial component that ties these dimensions together in successful virtual collaboration. Participants shared experiences that highlighted how familiarity with online collaboration tools directly impacts team communication and efficiency. One participant reflected on a challenge he encountered in virtual collaboration: "I didn't prepare my equipment in advance. Later, when joining the meeting, some issues occurred." Moreover, an in-depth understanding of the features offered by these tools is crucial for enhancing collaborative efforts. For example, one participant noted, "When we have an online meeting and need to set up an anonymous poll or Q&A session, it involves technical aspects of the meeting software. If you are not familiar with it, it can significantly affect efficiency." These examples emphasize the necessity of mastering tools and being technically prepared for smooth virtual collaboration.

**4.4 Interrelationship of behavioral indicators**

The findings of this study demonstrate that the aforementioned behavioral indicators are essential for virtual teamwork competency. However, their frequent demonstration alone does not ensure enhanced competency. Instead, the effectiveness of these behaviors also depends on their appropriate contextual application. Since these behavioral indicators are highly interconnected rather than functioning independently, their use in assessment requires careful consideration of their interrelationships. Through structured interviews, we systematically captured students' perceptions of these interconnections, providing a deeper understanding of their dynamics in virtual teamwork contexts.

**Interactions within the group dimension**

In the early stages of online collaboration, thorough task analysis is essential before labor division, which ensures that tasks are clearly defined and aligned with team members' strengths. One participant described their team's initial struggle: "We were all new and didn't understand the competition, so roles weren't clearly defined." In contrast, another participant shared their team's success story, where they "sat together to analyze the task and determine strategies," resulting in a more effective division of labor. Additionally, mutual understanding among team members is critical before dividing tasks. One participant described a chaotic experience: "Before, it was chaotic online, with one person handling too much instead of dividing tasks based on ability." By understanding each member's capabilities, teams can strategically align roles, maximizing strengths and maintaining a balanced workload. For instance, one participant attributed their success to this approach: "Our success lies in the reasonable division of work because we are from different specialties." This proactive strategy allows work to be "skillfully distributed," leveraging diverse expertise to enhance efficiency.

Mutual understanding also serves as the foundation for providing support and suggestions, as team members bring diverse personalities, work styles, and emotional needs. One participant highlighted the importance of adapting their approach: "There's no universal formula. Some people respond better to encouragement and gain a sense of accomplishment when praised." Similarly, offering suggestions requires sensitivity. As one participant noted, "If there's a problem, just say it directly. Most of us in STEM fields are logical thinkers, and even if the feedback is harsh, we understand it's for the betterment of the task." However, others preferred a more tactful approach, as shared by another participant: "Rather than raising suggestions during meetings, I'd prefer if someone gave me feedback privately—it's much easier to accept that way."

Furthermore, there is a significant relationship between information exchange and time management. Poor time management in online settings can hinder communication, as one participant observed: "I was free when he wasn't, and vice versa, which led to no communication or meetings at all." Another participant noted, "Poor time management restricts communication to images, videos, or text, which can often result in misunderstandings." To address this, one participant implemented a structured approach: "First, I understood their rest times and class schedules, and used those times to discuss work. We adopted a twice-weekly report schedule, once mid-week and once on the weekend." Proper time management ensures team members are available simultaneously, facilitating a timely and coherent flow of information.

The effectiveness of information exchange is also linked to the timeliness and positivity of feedback. Enthusiasm can significantly impact team dynamics, as one participant explained: "If everyone is proactive and enthusiastic, things would be much better. Sometimes, only one person is talking, and it gets awkward when others don't respond." Conversely, indifferent responses can stifle communication, as another participant shared: "No matter how much you communicate or carefully choose your words, some people still respond with a cold 'noted,' which creates an uninviting atmosphere."

Lastly, effective conflict resolution hinges on a keen awareness of the team's emotional dynamics. Participants highlighted the value of light-hearted approaches to defuse tension and maintain a positive atmosphere. For instance, one participant shared,

"When there are small conflicts and the atmosphere feels awkward, a touch of humor can work wonders to lighten things up—it really helps." Another echoed this sentiment, stating, "A well-timed joke can often resolve minor disagreements," underscoring the role of humor in easing friction. Additionally, a participant emphasized the importance of proactive emotional management: "When I notice team members feeling off, I make an effort to lighten the mood and restore a sense of harmony." These insights collectively highlight the significance of addressing emotional undercurrents, both before and after conflicts arise, to foster a collaborative and supportive team environment.

**Interactions between individual and group dimensions**

Labor division is not just about assigning tasks but also about proactively upholding team responsibility. Participants noted that a lack of initiative during task assignments often leads to disengagement. As one participant observed, "When tasks were assigned, everyone tended to avoid responsibility, saying things like 'I can't do this' or 'I've never done that before.'" To address this, some team members took the initiative to compensate for uneven workloads. For example, a participant shared, "If I felt I did less earlier, I would take on more during the final stages, like writing the summary document, to maintain team harmony." This sense of responsibility also plays a vital role in managing team emotions. When team morale is low, proactive actions, such as demonstrating determination to overcome challenges, can uplift the group. As one participant explained, "I hope to see actions that show we can succeed, like a determination to push through difficulties. Words of encouragement help, but actions inspire more."

Labor division is also closely tied to ensuring quality completion. Each team member should verify that assigned tasks are achievable. One participant emphasized the importance of this alignment: "After receiving my tasks, I realized the workload was too heavy for me as a beginner. I immediately discussed this with the team leader, who adjusted my responsibilities based on my capabilities." This proactive communication not only ensured quality work but also prevented delays caused by unspoken struggles. As another participant noted, "Some members pretend to have completed their tasks

when they haven't, which ultimately harms the team's efficiency."

In addition, progress reporting plays a pivotal role in ensuring on-time task completion. In online settings, where visibility into others' progress is limited, regular updates help maintain accountability. One participant admitted, "Without timely communication, I tend to procrastinate until the deadline." To address this, teams can establish regular check-ins. As another participant suggested, "In online collaboration, frequent communication is essential to track progress and keep everyone motivated."

## 5. Discussion

Based on the collected data, this study developed a behavior-oriented theoretical framework grounded in group dynamics theory to assess the virtual teamwork competency of engineering students. This framework consists of 15 behavioral indicators categorized into three dimensions: Group task dimension, Individual social dimension, and Individual task dimension, while also highlighting the interrelationships among these indicators. This structure reflects the complex interplay between individual and group dynamics as well as the social and task dimensions of teamwork. It aligns with the multi-level nature of teamwork competency described in prior research, which spans individual contributions, group interactions, and overarching task objectives (Murzi et al., 2020). Additionally, it resonates with Tuckman's (1965) observations on group development, emphasizing the dual processes of teams: achieving task completion while simultaneously fostering interpersonal relationships among members. These processes often overlap and mutually reinforce one another, further underscoring the interconnected nature of effective teamwork.

Within the three dimensions, some behavior-oriented indicators identified in this study also resonate with key themes in existing literature, such as task execution coordination and time management in the group task dimension. After task allocation, information sharing among members can enhance virtual team performance and facilitate team development processes (Algesheimer et al., 2011; A. Jony & Serradell Lopez, 2018). Moreover, as time coordination has been widely recognized as a significant challenge in asynchronous learning settings, Avery Gomez et al. (2009) emphasized that optimizing time allocation are vital for improving the efficiency of team-based learning in virtual environments, where misaligned schedules can frequently disrupt team progress. And in the individual social dimension, our theoretical framework highlights the importance of social cohesion, which refers to a connection and mutual attraction within the group that develops from the social relationships among group members (Carron et al., 1985; Seashore, 1954). The behavior indicators in our framework, such as mutual understanding, mutual support, emotional regulation, and active feedback, all contribute to fostering social cohesion. Lee et al. (2023) underlines the importance of emotional exchanges in achieving team objectives, noting that teams capable of recognizing and responding to the emotions of their members are better equipped to collaborate effectively. Besides, the role of feedback, as highlighted in this study, reinforces Hertel et al. (2005) argument that timely feedback bridges spatial and temporal gaps in virtual teams, enhancing both trust and cohesion. Conversely, previous research (e.g. Losada et al., 1990) indicates that the absence of process feedback may hinder relational exchanges, weakening the social bonds essential for effective collaboration. These interactions are considered key to building "swift" trust and are positively correlated with task cohesion, significantly contributing to team performance (Kahai & Cooper, 1999; Potter & Balthazard, 2002).

Our framework builds upon and extends existing models by incorporating indicators that are less commonly addressed in previous online collaboration frameworks. For instance, previous models have rarely mentioned indicators related to self-management. In our framework, the individual task dimension highlights the importance of personal accountability and technical competence in achieving effective online collaboration. This finding aligns with earlier empirical research on online collaboration: An et al. (2008) point out that in the context of online collaboration, where team members must often work independently without close supervision, self-discipline and technical readiness become even more critical. Furthermore, Avery Gomez et al. (2009) emphasize the need for online team members to be technically proficient, as technological failures or limitations can disrupt the collaboration process and reduce efficiency.

Additionally, in existing online collaboration frameworks, the significance of the interconnections between elements is often underemphasized. Our study identifies some interrelationships among behavioral indicators that strengthen effective teamwork, several of which are supported by previous research. For instance, the relationship between information exchange and active feedback identified in our study aligns with the findings of Hertel et al. (2005), who provide evidence that feedback on social processes helps bridge spatial disconnects in virtual teams, fostering cohesion and trust—both of which are directly tied to effective information exchange. This connection is further reinforced by Tseng et al. (2009), who highlight that establishing clear protocols for responding to communication, along with providing encouraging and timely feedback, creates a safe and supportive online learning environment. Similarly, Nakayama et al. (2018) support the connection between mutual understanding and labor division, demonstrating that interventions at the individual level can improve group coordination, resulting in a more spontaneous and effective division of labor. And the study by Ayoko & Konrad (2009) provides foundational insights into the connections between emotional dynamics and conflict resolution, suggesting that emotional awareness is crucial for effective conflict management and pointing to the need for explicit training and tools that help teams recognize and act on emotional cues proactively.

The interaction between individual and group dimensions is also noted in past literature. For example, Curşeu et al. (2008) highlight the importance of proactively upholding team responsibility to facilitate smoother labor division. This enables team members to dynamically adjust their contributions based on the evolving needs of the group, fostering a sense of shared responsibility. Conversely, when individuals fail to proactively uphold their responsibilities, it creates gaps in task execution, leading to inefficiencies and frustration among team members (Zhang et al., 2009). Additionally, the relationship between labor division and ensuring quality completion is further explored by Lin et al. (2014), who conclude through their analysis of novice Agile team members that optimal task allocations require aligning task complexity with the skill levels of the individuals involved.

**Implications**

This proposed framework offers a promising foundation for assessing virtual teamwork competency in the future. For students, it introduces behavioral benchmarks that serve as clear standards for evaluating their virtual teamwork competency. These benchmarks facilitate a structured self-assessment process, helping students identify their strengths and areas needing improvement. For educators, this framework can be first used for designing assessments that specifically target skills. It provides a clear structure to develop assessments that accurately reflect the competencies needed for effective online teamwork. Additionally, the framework offers concrete, observable metrics, enabling educators to use transcripts and direct observations for evaluation. As AI technology continues to advance, educators can further utilize technology to code behaviors and conduct detailed assessments, enhancing both objectivity and accuracy. The framework also facilitates the delivery of customized feedback, whether from educators or AI systems. Its behavior-oriented nature ensures that feedback is detailed and actionable, providing students with a clear understanding of their strengths and weaknesses and guiding them in effectively enhancing their teamwork competency.

Beyond its application in assessments, the framework also aids in the development of educational programs, workshops, and exercises aimed at enhancing online collaboration skills for engineering students. By focusing on behavior-oriented indicators, educators can design curriculum components that address key aspects of collaboration. This ensures that students engage with practical exercises that reflect real-world engineering challenges, bridging the gap between theory and practice.

## 6. Conclusion

By analyzing focus group interviews grounded in participants' lived experiences, this study developed a behavior-oriented framework for virtual teamwork competencies among engineering students, identifying three key dimensions along with their associated behavioral indicators and interrelationships. Compared to traditional forms of assessment evidence, behavioral indicators are relatively easier to observe, allowing for a more objective evaluation of teamwork competencies based on the collaborative process itself. Furthermore, the often-overlooked interconnections between these behavioral indicators offer a more comprehensive foundation for assessing teamwork performance. These findings address existing gaps in current frameworks and provide a more holistic perspective on virtual teamwork competency.

Despite its contributions, this study has certain limitations. The reliance on self-reported data collected through focus group interviews introduces potential bias, as participants may unintentionally omit or misrepresent behaviors. Additionally, while the qualitative approach provides deep insights into participants' experiences, it does not allow for quantifiable validation of the framework's indicators or interrelationships, limiting its applicability in broader contexts without further testing. Future research should prioritize validating the framework through empirical studies in diverse contexts. Another important direction is to investigate the effectiveness of using these indicators to evaluate team performance in complex virtual environments, potentially leveraging AI to monitor behaviors and provide deeper insights. Such research would provide stronger empirical evidence for the applicability of the framework and deepen our understanding of how these competencies influence team outcomes in real-world virtual collaboration settings.

# Reference

Aguado, D., Rico, R., Sánchez-Manzanares, M., & Salas, E. (2014). Teamwork Competency Test (TWCT): A step forward on measuring teamwork competencies. *Group Dynamics: Theory, Research, and Practice*, *18*(2), 101–121. https://doi.org/10.1037/a0036098

Algesheimer, R., Dholakia, U. M., & Gurău, C. (2011). Virtual Team Performance in a Highly Competitive Environment. *Group & Organization Management*, *36*(2), 161–190. https://doi.org/10.1177/1059601110391251

An, H., Kim, S., & Kim, B. (2008). Teacher Perspectives on Online Collaborative Learning: Factors Perceived as Facilitating and Impeding Successful Online Group Work. *Contemporary Issues in Technology and Teacher Education*, *8*(1), 65-.

Aritz, J., Walker, R., & Cardon, P. (2018). Media Use in Virtual Teams of Varying Levels of Coordination. *BUSINESS AND PROFESSIONAL COMMUNICATION QUARTERLY*, *81*(2), 222–243. https://doi.org/10.1177/2329490617723114

Avery Gomez, E., Wu, D., & Passerini, K. (2009). Traditional, Hybrid and Online Teamwork: Lessons from the Field. *Communications of the Association for Information Systems*, *25*, 395–412. https://doi.org/10.17705/1CAIS.02533

Avry, S., Chanel, G., Bétrancourt, M., & Molinari, G. (2020). Achievement appraisals, emotions and socio-cognitive processes: How they interplay in collaborative problem-solving? *Computers in Human Behavior*, *107*, 106267. https://doi.org/10.1016/j.chb.2020.106267

Ayoko, O. B., & Konrad, A. M. (2009). Process development in project teams and the emergence of team members' conflict and emotions in a virtual environment. *ANZAM 2009 Conference*, *Paper 460*, 1–16.

Beal, D. J., Cohen, R. R., Burke, M. J., & McLendon, C. L. (2003). Cohesion and Performance in Groups: A Meta-Analytic Clarification of Construct Relations. *Journal of Applied Psychology*, *88*(6), 989–1004. https://doi.org/10.1037/0021-9010.88.6.989

Brannick, M. T., Salas, E., & Prince, C. W. (1997). A Conceptual Framework for Teamwork Measurement. In *Team Performance Assessment and Measurement* (pp. 31–56). Psychology Press. https://doi.org/10.4324/9781410602053-9

Brewer, P. E., Mitchell, A., Sanders, R., Wallace, P., & Wood, D. D. (2015). Teaching and Learning in Cross-Disciplinary Virtual Teams. *IEEE Transactions on Professional Communication*, *58*(2), 208–229. IEEE Transactions on Professional Communication. https://doi.org/10.1109/TPC.2015.2429973

Britton, E., Simper, N., Leger, A., & Stephenson, J. (2017). Assessing teamwork in undergraduate education: A measurement tool to evaluate individual teamwork skills. *Assessment & Evaluation in Higher Education*, *42*(3), 378–397. https://doi.org/10.1080/02602938.2015.1116497

Carron, A. V., Widmeyer, W. N., & Brawley, L. R. (1985). The Development of an Instrument to Assess Cohesion in Sport Teams: The Group Environment Questionnaire. *Journal of Sport Psychology*, *7*(3), 244–266. https://doi.org/10.1123/jsp.7.3.244

Chan, C. K. Y. (2022). *Assessment for Experiential Learning*. Routledge. https://doi.org/10.4324/9781003018391

Chan, C. K. Y., & Yeung, N. C. J. (2020). Students' 'approach to develop' in holistic competency: An adaption of the 3P model. *Educational Psychology*, *40*(5), 622–642. https://doi.org/10.1080/01443410.2019.1648767

Chavis, S. E., Anagnostopoulos-King, F. V., Syme, S. L., Varlotta, S., Noonan, K. E., & Congdon, H. B. (2024). In-person to virtual interprofessional education: Teamwork attitudes and skills among dental and dental hygiene students. *Journal of Dental Education*, *88*(11), 1481–1489. https://doi.org/10.1002/jdd.13636

Chen, F., Sager, J., Corbitt, G., & Gardiner, S. C. (2008). Incorporating Virtual Teamwork Training into MIS Curricula. *Journal of Information Systems Education*, *19*(1), 29–41.

Chhabria, K., Black, E., Giordano, C., & Blue, A. (2019). Measuring health professions students' teamwork behavior using peer assessment: Validation of an online tool. *Journal of Interprofessional Education & Practice*, *16*, 100271. https://doi.org/10.1016/j.xjep.2019.100271

Curşeu, P. L., Schalk, R., & Wessel, I. (2008). How do virtual teams process information? A literature review and implications for management. *Journal of Managerial Psychology*, *23*(6), 628–652. https://doi.org/10.1108/02683940810894729

Dickinson, T. L., & McIntyre, R. M. (1997). A conceptual framework for teamwork measurement. In *Team performance assessment and measurement: Theory, methods, and applications* (pp. 19–43). Lawrence Erlbaum Associates Publishers. https://doi.org/10.4324/9781410602053

Falls, I., Bahhouth, V., Chuang, C. M., & Bahhouth, J. (2014). Factors Influencing Students' Perceptions of Online Teamwork. *Sage Open*, *4*(1), 2158244014525415. https://doi.org/10.1177/2158244014525415

Fathi, M., Ghobakhloo, M., & Syberfeldt, A. (2019). An Interpretive Structural Modeling of Teamwork Training in Higher Education. *Education Sciences*, *9*(1). https://doi.org/10.3390/educsci9010016

Flanagan, J. C. (1954). The critical incident technique. *Psychological Bulletin*, *51*(4), 327–358. https://doi.org/10.1037/h0061470

Forsyth, D. R. (2018). *Group Dynamics*. Cengage Learning.

García, M. G., López, C. B., Molina, E. C., Casas, E. E., & Morales, Y. A. R. (2016a). Development and evaluation of the team work skill in university contexts. Are virtual environments effective? *International Journal of Educational Technology in Higher Education*, *13*, 1–11. https://doi.org/10.1186/s41239-016-0014-1

Guenaga, M., Eguiluz, A., Rayon, A., Nunez, A., & Quevedo, E. (2014). A serious game to develop and assess teamwork competency. *2014 International Symposium on Computers in Education (SIIE)*, 183–188. https://doi.org/10.1109/SIIE.2014.7017727

Havyer, R. D. A., Wingo, M. T., Comfere, N. I., Nelson, D. R., Halvorsen, A. J., McDonald, F. S., & Reed, D. A. (2014). Teamwork Assessment in Internal Medicine: A Systematic Review of Validity Evidence and Outcomes. *Journal of General Internal Medicine : JGIM*, *29*(6), 894–910. https://doi.org/10.1007/s11606-013-2686-8

Hertel, G., Geister, S., & Konradt, U. (2005). Managing virtual teams: A review of current empirical research. *Human Resource Management Review*, *15*(1), 69–95. https://doi.org/10.1016/j.hrmr.2005.01.002

Hertel, G., Konradt, U., & Voss, K. (2006). Competencies for virtual teamwork: Developmentand validation of a web-based selection tool for members of distributed teams. *European Journal of Work and Organizational Psychology*, *15*(4), 477–504. https://doi.org/10.1080/13594320600908187

Hosseini, M. R., Martek, I., Chileshe, N., Zavadskas, E. K., & Arashpour, M. (2018). Assessing the Influence of Virtuality on the Effectiveness of Engineering Project Networks: "Big Five Theory" Perspective. *Journal of Construction Engineering and Management*, *144*(7), 04018059. https://doi.org/10.1061/(ASCE)CO.1943-7862.0001494

Hu, W., & Chan, C. K. Y. (2024). Evaluating technological interventions for developing teamwork competency in higher education: A systematic review and meta-ethnography. *Studies in Educational Evaluation*, *83*, 101382. https://doi.org/10.1016/j.stueduc.2024.101382

Ikonen, J., Knutas, A., Wu, Y., & Agudo, I. (2015). Is the world ready or do we need more tools for programming related teamwork? *Proceedings of the 15th Koli Calling Conference on Computing Education Research*, 33–39. https://doi.org/10.1145/2828959.2828978

Jony, A. I., & Serradell-López, E. (2021). An Evaluation of Virtual Teamwork Model in Online Higher Education. In A. Visvizi, M. D. Lytras, & N. R. Aljohani (Eds.), *Research and Innovation Forum 2020* (pp. 199–216). Springer International Publishing. https://doi.org/10.1007/978-3-030-62066-0_16

Jony, A., & Serradell Lopez, E. (2018). *Key Performance Indicators of Virtual Teamwork: A Conceptual Framework*. 5059–5068. https://doi.org/10.21125/iceri.2018.2153

Kahai, S. S., & Cooper, R. B. (1999). *The Effect of Computer-Mediated Communication on Agreement and Acceptance* (world). https://www.tandfonline.com/doi/pdf/10.1080/07421222.1999.11518238

Koh, E., Hong, H., & Tan, J. P.-L. (2018). Formatively assessing teamwork in technology-enabled twenty-first century classrooms: Exploratory findings of a teamwork awareness programme in Singapore. *Asia Pacific Journal of Education*, *38*(1), 129–144. https://doi.org/10.1080/02188791.2018.1423952

Kotlyar, I., Krasman, J., & Fiksenbaum, L. (2021). Virtual high-fidelity simulation assessment of teamwork skills: How do students react? *Journal of Research on Technology in Education*, *53*(3), 333–352. https://doi.org/10.1080/15391523.2020.1783401

Lacher, L. L., & Biehl, C. (2019). Investigating team effectiveness using Discord: A case study using a gaming collaboration tool for the CS classroom. *International Conference on Frontiers in Education: Computer Science and Computer Engineering, Las Vegas, USA*.

Lee, Y., Jung, J.-H., Kim, H., Jung, M., & Lee, S.-S. (2023). Comparative Case Study of Teamwork on Zoom and Gather.Town. *Sustainability*, *15*(2), Article 2. https://doi.org/10.3390/su15021629

Lepsinger, R., & DeRosa, D. (2010). *Virtual Team Success: A Practical Guide for Working and Leading from a Distance*. John Wiley & Sons.

Lewin, K. (1947). Frontiers in group dynamics: Concept, method and reality in social science; social equilibria and social change. *Human Relations*, *1*, 5–41. https://doi.org/10.1177/001872674700100103

Lin, J., Yu, H., Shen, Z., & Miao, C. (2014). Studying task allocation decisions of novice agile teams with data from agile project management tools. *Proceedings of the 29th ACM/IEEE International Conference on Automated Software Engineering*, 689–694. https://doi.org/10.1145/2642937.2642959

Linnes, C. (2020). Embracing the Challenges and Opportunities of Change Through Electronic Collaboration. *International Journal of Information Communication Technologies and Human Development*, *12*(4), 37–58. https://doi.org/10.4018/IJICTHD.20201001.oa1

Losada, M., Sanchez, P., & Noble, E. E. (1990). Collaborative technology and group process feedback: Their impact on interactive sequences in meetings. *Proceedings of the 1990 ACM Conference on Computer-Supported Cooperative Work*, 53–64. https://doi.org/10.1145/99332.99341

Marra, R. M., Steege, L., Tsai, C.-L., & Tang, N.-E. (2016). Beyond "group work": An integrated approach to support collaboration in engineering education. *International Journal of STEM Education*, *3*(1), Article 1. https://doi.org/10.1186/s40594-016-0050-3

Mayer, C., Sivatheerthan, T., Mütze-Niewöhner, S., & Nitsch, V. (2023). Sharing leadership behaviors in virtual teams: Effects of shared leadership behaviors on team member satisfaction and productivity. *Team Performance Management: An International Journal*, *29*(1/2), 90–112. https://doi.org/10.1108/TPM-07-2022-0054

McIntyre, R. M., Morgan Jr, B. B., Salas, E., & Glickman, A. S. (1988). Teamwork from team training: New evidence for the development of teamwork skills during operational training. *Proceedings of the 10th Annual Interservice/Industry Training Systems Conference*, 21–27.

Murzi, H., Chowdhury, T., Ek, J., & Ulloa, B. (2020). Working in Large Teams: Measuring the Impact of a Teamwork Model to Facilitate Teamwork Development in Engineering Students Working in a Real Project. *International Journal of Engineering Education*, *36*.

Nakayama, S., Diner, D., Holland, J., Bloch, G., Porfiri, M., & Nov, O. (2018). The Influence of Social Information and Self-expertise on Emergent Task Allocation in Virtual Groups. *Frontiers in Ecology and Evolution*, *6*. https://doi.org/10.3389/fevo.2018.00016

Petkovic, D., Thompson, G., Todtenhoefer, R., Huang, S., Levine, B., Parab, S., Singh, G., Soni, R., & Shrestha, S. (2010). Work in progress — e-TAT: Online tool for teamwork and "soft skills" assessment in software engineering education. *2010 IEEE Frontiers in Education Conference (FIE)*, S1G-1-S1G-3. https://doi.org/10.1109/FIE.2010.5673130

Piccoli, G., Powell, A., & Ives, B. (2004). Virtual teams: Team control structure, work processes, and team effectiveness. *Information Technology & People*, *17*(4), 359–379. https://doi.org/10.1108/09593840410570258

Porter, C. O. L. H., Hollenbeck, J. R., Ilgen, D. R., Ellis, A. P. J., West, B. J., & Moon, H. (2003). Backing up behaviors in teams: The role of personality and legitimacy of need. *Journal of Applied Psychology*, *88*(3), 391–403. https://doi.org/10.1037/0021-9010.88.3.391

Potter, R. E., & Balthazard, P. A. (2002). Understanding human interactions and performance in the virtual team. *JITTA : Journal of Information Technology Theory and Application*, *4*(1), 1–23.

Rousseau, V., Aubé, C., & Savoie, A. (2006). Teamwork Behaviors: A Review and an Integration of Frameworks. *Small Group Research*, *37*(5), 540–570. https://doi.org/10.1177/1046496406293125

Salas, E., Rosen, M. A., Burke, C. S., & Goodwin, G. F. (2009). The wisdom of collectives in organizations: An update of the teamwork competencies. In *Team effectiveness in complex organizations: Cross-disciplinary perspectives and approaches* (pp. 39–79). Routledge/Taylor & Francis Group.

Schulze, J., & Krumm, S. (2017). The "virtual team player": A review and initial model of knowledge, skills, abilities, and other characteristics for virtual collaboration. *Organizational Psychology Review*, *7*(1), 66–95. https://doi.org/10.1177/2041386616675522

Seashore, S. E. (1954). *Group cohesiveness in the industrial work group* (pp. vi, 107). Univ. of Michigan, Survey Research.

Siebdrat, F., Hoegl, M., & Ernst, H. (2009). How to Manage Virtual Teams. *MIT Sloan Management Review*, *50*(4), 63–68.

Strom, P. S., Strom, R. D., & Moore, E. G. (1999). Peer and self-evaluation of teamwork skills. *Journal of Adolescence*, *22*(4), 539–553. https://doi.org/10.1006/jado.1999.0247

Swartz, S., Barbosa, B., Crawford, I., & Luck, S. (2021). Minding the Competency Gap From College to Career: The Value of Virtual Teaming and VLEs for Skill Development. In *Developments in Virtual Learning Environments and the Global Workplace* (pp. 268–288). IGI Global. https://doi.org/10.4018/978-1-7998-7331-0.ch014

Tian, B., Zheng, Y., Zhuang, Z., Luo, H., Zhang, Y., & Wang, D. (2024). Group Haptic Collaboration: Evaluation of Teamwork Behavior during VR Four-Person Rowing Task. *IEEE Transactions on Haptics*, *17*(3), 384–395. IEEE Transactions on Haptics. https://doi.org/10.1109/TOH.2023.3346683

Tseng, H., Wang, C.-H., Ku, H.-Y., & Sun, L. (2009). Key factors in online collaboration and their relationship to teamwork satisfaction. Quarterly Review of Distance Education, 10(2). *Quarterly Review of Distance Education*, *10*, 195–206.

Tuckman, B. W. (1965). Developmental sequence in small groups. *Psychological Bulletin*, *63*(6), 384–399. https://doi.org/10.1037/h0022100

Vance, K., Kulturel-Konak, S., & Konak, A. (2015). Teamwork efficacy and attitude differences between online and face-to-face students. *2015 IEEE Integrated STEM Education Conference*, 246–251. https://doi.org/10.1109/ISECon.2015.7119933

Varela, O., & Mead, E. (2018). Teamwork skill assessment: Development of a measure for academia. *Journal of Education for Business*, *93*(4), 172–182. https://doi.org/10.1080/08832323.2018.1433124

Vivian, R., Falkner, K., Falkner, N., & Tarmazdi, H. (2016). A Method to Analyze Computer Science Students' Teamwork in Online Collaborative Learning Environments. *ACM Transactions on Computing Education*, *16*(2), 7:1-7:28. https://doi.org/10.1145/2793507

Wei, S., Tan, L., Zhang, Y., & Ohland, M. (2024). The effect of the emergency shift to virtual instruction on student team dynamics. *European Journal of Engineering Education*, *49*(1), 139–163. https://doi.org/10.1080/03043797.2023.2217422

Willox, S., Morin, J., & Avila, S. (2022). Benefits of individual preparation for team success: Planning for virtual team communication, conflict resolution and belonging. *Team Performance Management: An International Journal*, *29*(1/2), 1–14. https://doi.org/10.1108/TPM-03-2022-0022

Yoon & Seung-Won. (2003). *Examination of member behaviors, group processes, and development-shaping forces of virtual learning teams /*.

Zalesny, M. D. (1990). Rater confidence and social influence in performance appraisals. *Journal of Applied Psychology*, *75*(3), 274–289. https://doi.org/10.1037/0021-9010.75.3.274

Zhang, J., Scardamalia, M., Reeve, R., & Messina, R. (2009). Designs for Collective Cognitive Responsibility in Knowledge-Building Communities. *Journal of the Learning Sciences*, *18*(1), 7–44. https://doi.org/10.1080/10508400802581676